\documentclass[sigconf]{acmart}

\usepackage[font=small]{caption}
\usepackage[font=small]{subcaption}
\usepackage{amsmath}
\usepackage{gensymb}

\usepackage{wrapfig}
\usepackage[keeplastbox]{flushend}

\usepackage[normalem]{ulem}
\usepackage{blindtext}

\usepackage{booktabs}
\usepackage{tabularx}
\newcolumntype{b}{X}
\newcolumntype{s}{>{\hsize=.4\hsize}X}

\usepackage[inline]{enumitem}
\usepackage{xspace}

\usepackage{dblfloatfix}    
\usepackage{soul}

\newlist{romaninline}{enumerate*}{1}
\setlist[romaninline]{label=(\roman*)}

\usepackage[acronyms,nonumberlist,nopostdot,nomain,nogroupskip]{glossaries}
\newacronym{quic}{QUIC}{Quick UDP Internet Connections}
\newacronym{3gpp}{3GPP}{3rd Generation Partnership Project}
\newacronym{adc}{ADC}{Analog to Digital Converter}
\newacronym{5g}{5G}{5th Generation}
\newacronym{aimd}{AIMD}{Additive Increase Multiplicative Decrease}
\newacronym{am}{AM}{Acknowledged Mode}
\newacronym{amc}{AMC}{Adaptive Modulation and Coding}
\newacronym{aqm}{AQM}{Active Queue Management}
\newacronym{awgn}{AGWN}{Additive White Gaussian Noise}
\newacronym{balia}{BALIA}{Balanced Link Adaptation}
\newacronym{bdp}{BDP}{Bandwidth-Delay Product}
\newacronym{bf}{BF}{Beamforming}
\newacronym{cc}{CC}{Congestion Control}
\newacronym{cdf}{CDF}{Cumulative Distribution Function}
\newacronym{ci}{CI}{Close-in free space reference}
\newacronym{cn}{CN}{Core Network}
\newacronym{cqi}{CQI}{Channel Quality Information}
\newacronym{cp}{CP}{Control Plane}
\newacronym{csirs}{CSI-RS}{Channel State Information - Reference Signal}
\newacronym{dc}{DC}{Dual Connectivity}
\newacronym{dce}{DCE}{Direct Code Execution}
\newacronym{dci}{DCI}{Downlink Control Information}
\newacronym{dl}{DL}{Downlink}
\newacronym{dmr}{DMR}{Deadline Miss Ratio}
\newacronym{dmrs}{DMRS}{DeModulation Reference Signal}
\newacronym{e2e}{E2E}{End-to-End}
\newacronym{ecn}{ECN}{Explicit Congestion Notification}
\newacronym{edf}{EDF}{Earliest Deadline First}
\newacronym{enb}{eNB}{evolved Node Base}
\newacronym{epc}{EPC}{Evolved Packet Core}
\newacronym{es}{ES}{Edge Server}
\newacronym{fdma}{FDMA}{Frequency Division Multiple Access}
\newacronym{fdd}{FDD}{Frequency Division Duplexing}
\newacronym[firstplural=Radio Access Technologies (RATs)]{rat}{RAT}{Radio Access Technology}
\newacronym{fs}{FS}{Fast Switching}
\newacronym{ftp}{FTP}{File Transfer Protocol}
\newacronym{gnb}{gNB}{Next Generation Node Base}
\newacronym{harq}{HARQ}{Hybrid Automatic Repeat reQuest}
\newacronym{hetnet}{HetNet}{Heterogeneous Network}
\newacronym{hh}{HH}{Hard Handover}
\newacronym{hol}{HOL}{Head-of-Line}
\newacronym{ia}{IA}{Initial Access}
\newacronym{imt}{IMT}{International Mobile Telecommunication}
\newacronym{iot}{IoT}{Internet of Things}
\newacronym{los}{LOS}{Line of Sight}
\newacronym{lte}{LTE}{Long Term Evolution}
\newacronym{m2m}{M2M}{Machine to Machine}
\newacronym{mac}{MAC}{Medium Access Control}
\newacronym{mc}{MC}{Multi-Connectivity}
\newacronym{mcs}{MCS}{Modulation and Coding Scheme}
\newacronym{mec}{MEC}{Mobile Edge Cloud}
\newacronym{mi}{MI}{Mutual Information}
\newacronym{mimo}{MIMO}{Multiple Input, Multiple Output}
\newacronym{mmwave}{mmWave}{millimeter wave}
\newacronym{mr}{MR}{Maximum Rate}
\newacronym{mss}{MSS}{Maximum Segment Size}
\newacronym{mtd}{MTD}{Machine-Type Device}
\newacronym{mtu}{MTU}{Maximum Transmission Unit}
\newacronym{nsf}{NSF}{National Science Foundation}
\newacronym{nfv}{NFV}{Network Function Virtualization}
\newacronym{nlos}{NLOS}{Non Line of Sight}
\newacronym{nr}{NR}{New Radio}
\newacronym{ofdm}{OFDM}{Orthogonal Frequency Division Multiplexing}
\newacronym{pdcch}{PDCCH}{Physical Downlonk Control Channel}
\newacronym{pdcp}{PDCP}{Packet Data Convergence Protocol}
\newacronym{pdsch}{PDSCH}{Physical Downlink Shared Channel}
\newacronym{pdu}{PDU}{Packet Data Unit}
\newacronym{pf}{PF}{Proportional Fair}
\newacronym{pgw}{PGW}{Packet Gateway}
\newacronym{phy}{PHY}{Physical}
\newacronym{pbch}{PBCH}{Physical Broadcast Channel}
\newacronym[plural=\gls{mme}s,firstplural=Mobility Management Entities (MMEs)]{mme}{MME}{Mobility Management Entity}
\newacronym{prb}{PRB}{Physical Resource Block}
\newacronym{pss}{PSS}{Primary Synchronization Signal}
\newacronym{pucch}{PUCCH}{Physical Uplink Control Channel}
\newacronym{pusch}{PUSCH}{Physical Uplink Shared Channel}
\newacronym{rach}{RACH}{Random Access Channel}
\newacronym{ran}{RAN}{Radio Access Network}
\newacronym{red}{RED}{Random Early Detection}
\newacronym{rf}{RF}{Radio Frequency}
\newacronym{rlc}{RLC}{Radio Link Control}
\newacronym{rlf}{RLF}{Radio Link Failure}
\newacronym{rrc}{RRC}{Radio Resource Control}
\newacronym{rrm}{RRM}{Radio Resource Management}
\newacronym{rr}{RR}{Round Robin}
\newacronym{rs}{RS}{Remote Server}
\newacronym{rsrp}{RSRP}{Reference Signal Received Power}
\newacronym{rss}{RSS}{Received Signal Strength}
\newacronym{rtt}{RTT}{Round Trip Time}
\newacronym{rw}{RW}{Receive Window}
\newacronym{rx}{RX}{Receiver}
\newacronym{sa}{SA}{standalone}
\newacronym{sack}{SACK}{Selective Acknowledgment}
\newacronym{sap}{SAP}{Service Access Point}
\newacronym{sch}{SCH}{Secondary Cell Handover}
\newacronym{scoot}{SCOOT}{Split Cycle Offset Optimization Technique}
\newacronym{sdma}{SDMA}{Spatial Division Multiple Access}
\newacronym{sinr}{SINR}{Signal to Interference plus Noise Ratio}
\newacronym{sm}{SM}{Saturation Mode}
\newacronym{snr}{SNR}{Signal to Noise Ratio}
\newacronym{son}{SON}{Self-Organizing Network}
\newacronym{ss}{SS}{Synchronization Signal}
\newacronym{srs}{SRS}{Sounding Reference Signal}
\newacronym{sss}{SSS}{Secondary Synchronization Signal}
\newacronym{tb}{TB}{Transport Block}
\newacronym{tcp}{TCP}{Transmission Control Protocol}
\newacronym{tdd}{TDD}{Time Division Duplexing}
\newacronym{tdma}{TDMA}{Time Division Multiple Access}
\newacronym{tfl}{TfL}{Transport for London}
\newacronym{thz}{THz}{Terahertz}
\newacronym{tm}{TM}{Transparent Mode}
\newacronym{trp}{TRP}{Transmitter Receiver Pair}
\newacronym{tti}{TTI}{Transmission Time Interval}
\newacronym{ttt}{TTT}{Time-to-Trigger}
\newacronym{tx}{TX}{Transmitter}
\newacronym{ue}{UE}{User Equipment}
\newacronym{ul}{UL}{Uplink}
\newacronym{uml}{UML}{Unified Modeling Language}
\newacronym{um}{UM}{Unacknowledged Mode}
\newacronym{utc}{UTC}{Urban Traffic Control}
\newacronym{vm}{VM}{Virtual Machine}
\newacronym{rsrq}{RSRQ}{Reference Signal Received Quality}
\newacronym{rssi}{RSSI}{Received Signal Strength Indicator}
\newacronym{crs}{CRS}{Cell Reference Signal}
\newacronym{comp}{CoMP}{Coordinated Multi-Point}
\newacronym{cran}{C-RAN}{Cloud \acrlong{ran}}
\newacronym{ca}{CA}{Carrier Aggregation}
\newacronym{cco}{CC}{Carrier Component}
\newacronym{nsa}{NSA}{Non Stand Alone}
\newacronym{embb}{eMBB}{Enhanced Mobility Broadband}
\newacronym{bsr}{BSR}{Buffer Status Report}
\newacronym{srb}{SRB}{Service Radio Bearer}
\newacronym{scm}{SCM}{Spatial Channel Model}
\newacronym{sctp}{SCTP}{Stream Control Transmission Protocol}
\newacronym{mptcp}{MPTCP}{Multi-path TCP}
\newacronym{ietf}{IETF}{Internet Engineering Task Force}
\newacronym{os}{OS}{Operating System}
\newacronym{tls}{TLS}{Transport Layer Security}
\newacronym{rfc}{RFC}{Request for Comments}
\newacronym{http}{HTTP}{HyperText Transfer Protocol}
\newacronym{nat}{NAT}{Network Address Translation}
\newacronym{api}{API}{Application Programming Interface}
\newacronym{rto}{RTO}{Retransmission Timeout}
\newacronym{psc}{PSC}{Public Safety Communication}
\newacronym{rpgm}{RPGM}{Reference Point Group Mobility}
\newacronym{ic}{IC}{Incident Command}
\newacronym{rsu}{RSU}{Road Side Unit}
\newacronym{uav}{UAV}{Unmanned Aerial Vehicle}
\newacronym{usa}{U.S.}{United States}
\newacronym{vr}{VR}{Virtual Reality}
\newacronym{iab}{IAB}{Integrated Access and Backhaul}
\newacronym{wlan}{WLAN}{Wireless Local Area Network}
\newacronym{cots}{COTS}{Commercial Off-the-Shelf}
\newacronym{fpga}{FPGA}{Field Programmable Gate Array}
\newacronym{rcn}{RCN}{Research Coordination Network}
\newacronym{abg}{ABG}{Alpha-Beta-Gamma}
\newacronym{fi}{FI}{Floating Intercept}
\newacronym{uas}{UAS}{Unmanned Aerial System}
\newacronym{gps}{GPS}{Global Positioning System}
\newacronym{a2g}{A2G}{air-to-ground}
\newacronym{a2a}{A2A}{air-to-air}
\newacronym{uma}{UMa}{Urban Macro}
\newacronym{umi}{UMi}{Urban Micro}
\newacronym{rma}{RMa}{Rural Macro}
\newacronym{inoo}{InOo}{Indoor Open Office}
\newacronym{ple}{PLE}{path loss exponent}
\newacronym{aoa}{AoA}{Angle of Arrival}
\newacronym{aod}{AoD}{Angle of Departure}
\newacronym{toa}{ToA}{Time of Arrival}
\newacronym{mpc}{MPC}{Multi-path Component}
\newacronym{cir}{CIR}{Channel Impulse Response}
\newacronym{rt}{RT}{Ray-tracing}
\newacronym{tc}{TC}{Time Cluster}
\newacronym{sl}{SL}{Spatial Lobe}
\newacronym{6g}{6G}{Sixth Generation}
\newacronym{ns3}{ns-3}{Network Simulator 3}
\newacronym{fsc}{FS}{Fully Stochastic}
\newacronym{hbc}{HB}{Hybrid}
\newacronym{hpbw}{HPBW}{Half Power Beamwidth}

\title{Terahertz Communications Can Work in Rain and Snow:\\Impact of Adverse Weather Conditions on Channels at 140 GHz}

\author{Priyangshu Sen, Jacob Hall, Michele Polese, Vitaly Petrov, Duschia Bodet,\\Francesco Restuccia, Tommaso Melodia, Josep M. Jornet}
\affiliation{
\institution{Northeastern University, Boston, MA}}
\email{j.jornet@northeastern.edu}

\begin{abstract}
    Next-generation wireless networks will leverage the spectrum above 100 GHz to enable ultra-high data rate communications over multi-GHz-wide bandwidths. The propagation environment at such high frequencies, however, introduces challenges throughout the whole protocol stack design, from physical layer signal processing to application design. Therefore, it is fundamental to develop a holistic understanding of the channel propagation and fading characteristics over realistic deployment scenarios and ultra-wide bands. In this paper, we conduct an extensive measurement campaign to evaluate the impact of weather conditions on a wireless link in the 130-150 GHz band through a channel sounding campaign with clear weather, rain, and snow in a typical urban backhaul scenario. We present a novel channel sounder design that captures signals with -82 dBm sensitivity and 20~GHz of bandwidth. We analyze link budget, capacity, as well as channel parameters such as the delay spread and the K-factor. Our experimental results indicate that in the considered context the adverse weather does not interrupt the link, but introduces some additional constraints (e.g., high delay spread and increase in path loss in snow conditions) that need to be accounted for in the design of reliable \gls{6g} communication links above 100 GHz.
\end{abstract}



\copyrightyear{2022}
\acmYear{2022}
\setcopyright{acmlicensed}\acmConference[mmNets'22]{6th ACM Workshop on Millimeter-Wave and Terahertz Networks and Sensing Systems }{October 17, 2022}{Sydney, NSW, Australia}
\acmBooktitle{6th ACM Workshop on Millimeter-Wave and Terahertz Networks and Sensing Systems (mmNets'22), October 17, 2022, Sydney, NSW, Australia} \acmPrice{15.00}
\acmDOI{10.1145/3555077.3556470}
\acmISBN{978-1-4503-9509-0/22/10}

\begin{document}

\fancyhead{}

\maketitle

\glsresetall

\section{Introduction}
\label{sec:intro}
 
New wireless applications such as highly-immersive multimedia applications (e.g., AR/VR, 3D telepresence, the metaverse), require constantly increasing data rates in mobile scenarios. Other use cases include wireless backhauling for ultra-dense wireless access networks and non-terrestrial communications, among others. This calls for the development and deployment of new wireless technologies, with the goal to reach hundreds of Gigabit-per-second (Gbps) or Terabit-per-second (Tbps) data rates on the air interface.

\begin{figure*}[h!]
\centering
\includegraphics[width=0.9\linewidth]{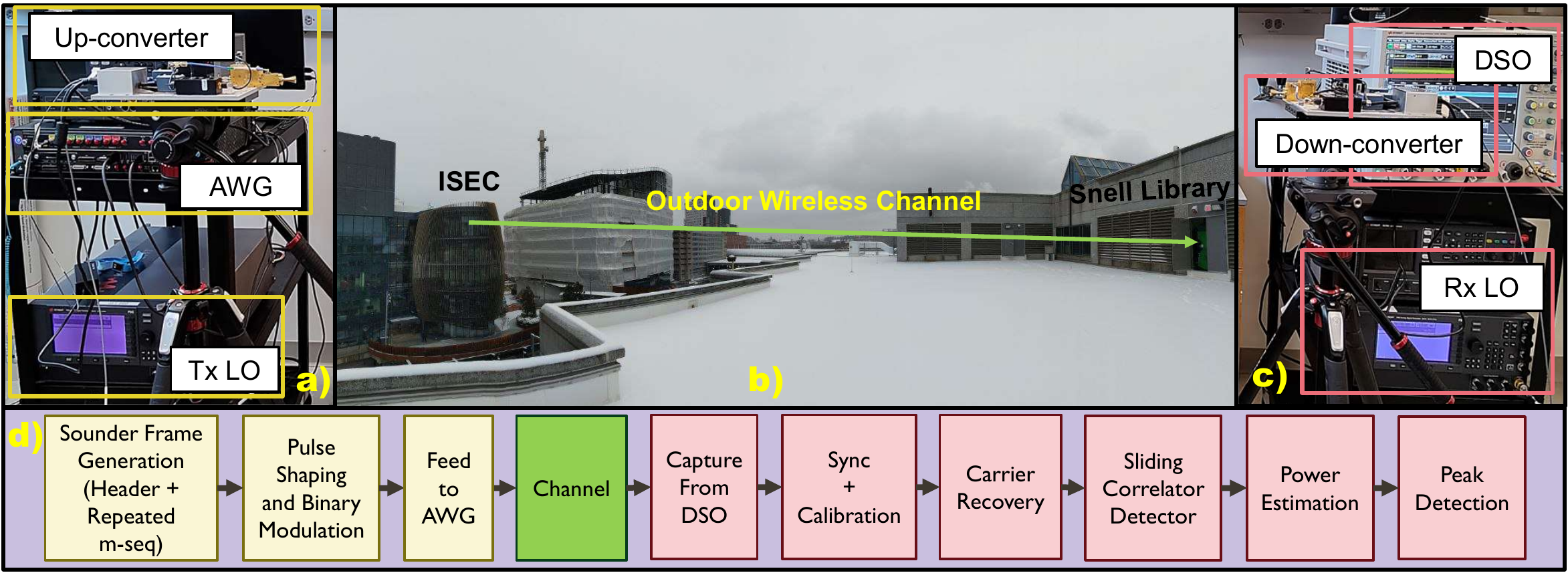}
\vskip -2ex
\caption{The NU Channel Sounder System. a) Transmitter Hardware. b) Outdoor Link. c) Receiver Hardware. d) Signal Processing Backend.
\vspace{-1.2em}
}
\label{fig:channel_sounder}
\end{figure*}

While \gls{5g} systems are now being deployed in the sub-6 GHz and lower \gls{mmwave} bands, the wireless research and industrial communities are now looking to \gls{6g} technologies~\cite{giordani2020toward}, candidate frequency bands~\cite{akyildiz2022terahertz}, and standardization~\cite{redmill20216g} to satisfy the need for higher data rates. In this regard, \gls{6g} networks will include communications in the spectrum above 100 GHz---the lower part of the terahertz bands---which has the potential to provide unparalleled communication data rates~\cite{akyildiz2022terahertz}.
This is possible thanks to advances in multiple areas around wireless systems, from devices and \gls{rf} components~\cite{singh2020design}, to physical and \gls{mac} layer design~\cite{polese2022dynamic}, and network and applications~\cite{polese2020toward}.

Establishing reliable data links at frequencies greater than 100 GHz  comes with multiple challenges. These are mostly related to the inherently harsh propagation environment due to the impact of molecular absorption in specific sub-bands above 100 GHz, the increased spreading loss, and the high sensitivity of such signals to blockage~\cite{akyildiz2022terahertz,moltchanov2022blockage}. These can be partially mitigated by exploiting high-gain directional antennas. At the same time, while directionality is fundamental to improving the link budget, it also introduces complex \gls{mac} and networking procedures to ensure different endpoints steer toward each other while trying to communicate~\cite{petrov2020micro}. These challenges---combined with the need to support extremely high data rates---impact the full protocol stack, with consequences on routing, transport layer, and application design, among others~\cite{polese2020toward}.

These challenges clearly highlight that a proper understanding of the wireless propagation and fading above 100 GHz will play a fundamental role in the design of robust, effective, and reliable communication links at these frequencies. Channel modeling above 100 GHz has spurred, physics-based research contributions~\cite{jornet2011channel}. More recently, we have also seen measurement-based campaigns through channel sounding techniques described in~\cite{chen2021channel,Statchannel}. Additionally, when considering realistic field deployments of \gls{6g} networks in urban and rural contexts, it is important to properly characterize the impact of adverse weather conditions (e.g., rain and snow) on the channel metrics and link performance. So far, the only experimental studies with snow focus just on the received power~\cite{mittleman2018snow,moon2015long} and thus do not provide a full-fledged channel characterization of the impact of adverse weather at these frequency bands. 

In this paper, we perform the first ultra-broadband, long-distance channel sounding campaign in the 130-150 GHz band with different weather conditions (clear, rain, and snow) in a typical urban backhaul scenario.
For this purpose, we have developed a fully tailored and adaptable a sliding-correlator-type channel sounder. This is capable of capturing multipath profiles with high resolution (0.1~ns) and dynamic range (60~dB) to describe the ultra-broadband nature of the link. 
We analyze the collected channel sounding data to provide an extensive evaluation of the weather impact on several parameters typically used in three-dimensional (3D) \gls{scm}, including path loss, K-factor, and delay spread. We validate the measurements by comparing the experimental results with physics-driven models. Finally, we analyze the capacity of a wireless link deployed in the same weather conditions.

Our results show that the impact of weather is limited and does not disrupt the wireless link in the urban context where we deployed our transceivers. The snow introduces the highest path loss increase -- around 13 dB when compared to clear weather -- while rain only increases loss by 1.3 dB. In addition, the delay spread with snow exceeds 0.5 ns for more than 10\% of the measured frames, which is enough to create inter-symbol interference (ISI) when communicating with 4 GHz of bandwidth or more. This analysis provides insights and takeaways that will be pivotal for the design of reliable communications and link layer solutions above 100 GHz.

The rest of the paper is organized as follows. Sec.~\ref{sec:measurement_setup} describes the channel sounder and the experiment's environment and methodology, while Sec.~\ref{sec:Exp_result} analyzes the experimental results, from link budget to channel and capacity metrics. Sec.~\ref{sec:conclusion} concludes the paper.

\section{Measurement Setup}
\label{sec:measurement_setup}

We first introduce the channel sounder developed as part of this effort, then describe our sounding methodology and scenario.

\subsection{NU Channel Sounder}
\label{subsec:Channel_sounder}

We developed a spread spectrum-based sliding correlator type channel sounder, the ``NU Channel Sounder'' named after Northeastern University (NU).
It is capable of capturing multipath profiles with high resolution and dynamic range, and is thus tailored to describe the ultra-broadband nature of the band above 100 GHz.
The system is divided into two parts – 1) Terahertz sounder frontends and 2) Signal processing backend.

\vspace{-.1cm}
\subsubsection{Terahertz Sounder frontends}
\label{subsubsec:Frontend}

The key components of the channel sounder frontends are developed based upon the ``TeraNova'' testbed~\cite{sen2021versatile,jornet2022ultra}. The transmitter and receiver components are shown in Fig.~\ref{fig:channel_sounder}~(a) and (c), respectively.
The system hardware at the transmitter is based on an arbitrary waveform generator (AWG), which can generate baseband (BB)/intermediate frequency (IF) channel sounding signals with up to 32 GHz of bandwidth. 
The BB/IF signal is fed to our THz RF frontend that up-converts the input signal with a mixer, driven by a multiplied local oscillator (LO) at 120~GHz, to produce a pass-band signal centered at 140~GHz.
The signal is received by another RF frontend that down-converts the pass-band signal in a similar manner.
The received BB/IF signal is fed to our digital storage oscilloscope (DSO), which can capture and store signals with 63.2~GHz of operational bandwidth. The saved signals are processed offline by the signal processing backend.

\vspace{-.1cm}
\subsubsection{Signal Processing Backend}
\label{subsubsec:Backend}

The signal processing backend is divided into transmitter and receiver sections, and its block diagram is shown in Fig.~\ref{fig:channel_sounder}~(d).
The backend defines the sounding signal and generates the digital BB/IF signal at the transmitter.
At the receiver, the backend processes the received signal to acquire the channel impulse response or multipath profile.
We use MATLAB to develop the different signal processing blocks of the backend.

The transmitted frame starts with an 8191-chips-long m-sequence header used for frame synchronization, calibration, and carrier recovery. This is followed by a 4095-chips-long m-sequence repeated 16 times that is used for channel sensing.
The full frame is modulated with binary phase-shift keying (BPSK) and pulse shaped with the root raised cosine filter to reduce out of the band leakage. The digital BB/IF signal is actualized by the AWG for further frequency up-conversion and transmission at 130-150 GHz band.

At the receiver, the pass-band signal is down-converted back to the BB/IF band and captured by the DSO for further processing. The header is used to synchronize with the start of a frame. Thereafter, the received signal is calibrated to eliminate the frequency selective response of THz frontend devices and connectors. The measured frequency response of each component is estimated as
\begin{equation}\label{eq:calibration}
H(k)=\biggl(\frac{P_r (k)-P_n}{P_s (k)}\biggr)^{1/2},
\end{equation}
where $P_r(k)$ is the received signal power with noise at the $k_{th}$ frequency, $P_s(k)$ refers to the transmitted signal power, and $P_n$ is noise power for the entire observation bandwidth. The inverse of normalized $H$ calibrates the received signals to mitigate the frequency selectivity of the hardware.
However, the utilization of this approach increases the noise floor by approximately 10~dB.

To synchronize with the IF carrier of the received signal, a dual phase lock loop (PLL) is used. The frequency offset is estimated by the slope of changing/sliding phase of the first PLL. Meanwhile, the phase offset is estimated from the output of the second PLL, which has feedback from the first PLL to eliminate phase sliding. To reduce the phase noise, an average of 16 readings is considered. 

The received BB/IF signal is correlated with a locally generated 4095-chips-long m-seq based BB/IF signal to extract the channel impulse response. The impulse response is averaged over 16 such repetitions to reduce the noise floor and falsely detected impulses. The actual received power of the multipath profile is obtained by normalizing the processed channel impulse response with the pulse energy and m-seq length~(4095). The peaks and corresponding delays are measured through the global convergence method by setting the threshold based on the dynamic range of the channel sounder. 

The channel sounder is capable of sounding 20~GHz of RF bandwidth (i.e., the operational bandwidth of the RF frontends' mixers) and can measure multipath components with 0.1~ns time resolution. Further, the sounder system can estimate the multipath channel response up to 130~dB of path loss with a substantial dynamic range of 60~dB (with 38~dBi antenna at both transceiver sides). Also, the sounder can detect the channel response for paths experiencing up to a 160~dB loss for line-of-sight (LoS) point-to-point link scenario by compromising over dynamic range. It can be used for Doppler measurements less than 135~KHz with 7.38~$\mu$s of measurement time.

\subsection{Sounding Environment and Methodology}
\label{subsec:sounding_env}

The measurement campaign is performed in an outdoor urban environment considering a static point-to-point link. The 70~m-long link is set up between two buildings on the Northeastern University campus.
We transmit from the 6$^{th}$ floor of the Interdisciplinary Science and Engineering Complex (ISEC) to the roof of the Snell Library, shown in Fig.~\ref{fig:channel_sounder}~(b).
The channel measurement was performed in three different weather conditions: (i)~Clear, (ii)~Rain, and (iii)~Snow, to characterize link performance in adverse weather. 
We recorded data in 3-minute intervals over 6 hours, collecting 120 frames of data for each scenario.
We use ``Accuweather'' weather 
service
to record real-time atmospheric conditions that are later used to estimate the theoretical link characteristics.

\section{Experimental Results}
\label{sec:Exp_result}

In our numerical study, we first analyze the link budget with considerations for adverse weather conditions and compare the empirical results to the results derived with analytical models in Sec.~\ref{subsec:Link_Budget_Analysis}. We then discuss the channel capacity values achievable in different setups in Sec.~\ref{subsec:Capacity_analysis}. We finally move into the findings of our sounding system with the detected multipath profiles and statistically determined channel metrics in Sec.~\ref{subsec:Multipath_profile} and Sec.~\ref{subsec:metrics}, respectively.

\subsection{Link Budget Analysis}
\label{subsec:Link_Budget_Analysis}
The link budget incorporates the transmit power, the hardware's gains and losses, the free-space path loss (FSPL), and the channel attenuation. The channel loss comes from molecular absorption, as well as scattering for the rain and snow scenarios.

\subsubsection{Link Budget}
\label{subsubsec:Link_Budget}

The theoretical power received, $P_{r,T}$, is modeled with our link budget
\begin{equation}\label{eq:link_budget}
    P_{r,T} = P_s + G_s + G_r + G_{h} - L_{h} - L_{FSPL} - L_{abs} - L_{sc}\ ,
\end{equation}
where $P_s$ is the transmit power before the antenna, $G_s,\ G_r$ are the transmitter and receiver antenna gains, $G_{h}$ is the receiver-side hardware gain, $L_{h}$ is the receiver-side hardware loss, $L_{FSPL}$ is the FSPL of the center frequency 140~GHz, $L_{abs}$ is the molecular absorption loss, and $L_{sc}$ is the loss caused by scattering.

The RF transmitter outputs $P_s$~=~16~dBm, which combined with $G_s$~=~38~dBi from the horn antenna produces an effective isotropic radiated power (EIRP) of approximately 54~dBm. A similar 38~dBi antenna is used at the receiver.
The hardware gain comes from the low noise amplifier (LNA) operating at IF, and the losses come from the mixer conversion loss and the cable insertion loss.

\subsubsection{Molecular Absorption and Rain Scattering}
The International Telecommunication Union (ITU) provides models for molecular absorption and rain scattering in their recommendations ITU-R P.676-12 \cite{iturp676} and ITU-R P.838-3 \cite{iturp838}, respectively.
These calculate loss for frequencies of 1-1000~GHz as functions of temperature, pressure, water vapor, and rainfall rate. The average observed values of the relevant atmospheric readings are given in Table~\ref{tab:weather_stats}.

\begin{table}[h]
\centering
\vspace{-1em}
\begin{tabular}{ccccc}
\multicolumn{1}{l}{}                & \textbf{\begin{tabular}[c]{@{}c@{}}Temp.\\ (C)\end{tabular}} & \textbf{\begin{tabular}[c]{@{}c@{}}Water Vapor\\ (g/m$^3$)\end{tabular}} & \textbf{\begin{tabular}[c]{@{}c@{}}Pressure\\ (hPa)\end{tabular}} & \textbf{\begin{tabular}[c]{@{}c@{}}Precip.\\ (mm/hr)\end{tabular}} \\ \cline{2-5} 
\multicolumn{1}{c|}{\textbf{Clear}} & \multicolumn{1}{c|}{12.23,$\ $ 0.60}                             & \multicolumn{1}{c|}{5.15,$\ $ 0.18}                                          & \multicolumn{1}{c|}{1026,$\ $ 0.51}                                & \multicolumn{1}{c|}{0,$\ $ 0}                                          \\ \cline{2-5} 
\multicolumn{1}{c|}{\textbf{Rain}}  & \multicolumn{1}{c|}{$\ $7.23,$\ $ 0.39}                              & \multicolumn{1}{c|}{7.09,$\ $ 0.05}                                          & \multicolumn{1}{c|}{1014,$\ $ 0.43}                                & \multicolumn{1}{c|}{1.84,$\ $ 1.22}                                    \\ \cline{2-5} 
\multicolumn{1}{c|}{\textbf{Snow}}  & \multicolumn{1}{c|}{-2.32,$\ $ 0.50}                             & \multicolumn{1}{c|}{3.54,$\ $ 0.03}                                          & \multicolumn{1}{c|}{1020,$\ $ 3.30}                                & \multicolumn{1}{c|}{0.45,$\ $ 0.08}                                    \\ \cline{2-5} 
\end{tabular}
\setlength\abovecaptionskip{-.3cm}
\caption{Mean and variance (left and right values of each cell) of weather data collected from "Accuweather" weather service.
\vspace{-1em}
}
\label{tab:weather_stats}
\end{table}
\noindent The scattering loss from rain, $L_{rain}$, is computed as
\begin{equation}\label{eq:rain_scattering}
    L_{rain} = kR_{rain}^\alpha\ ,
\end{equation}
where $R_{rain}$ is the rain precipitation rate in mm/hr, while the coefficients $k$ and $\alpha$ are calculated for horizontal polarization~\cite{iturp838}.

For the 130-150~GHz band and a link distance of 70~m, the loss from molecular absorption and rain scattering were in the order of 0.01~dB and 0.1~dB, respectively, thus ultimately negligible.

\subsubsection{Snow Scattering}
\label{subsubsec:snow_scattering}

Although there is no standard model for snow scattering of terahertz waves, pre-existing literature that can be leveraged  \cite{mittleman2018snow,federici2019snow}, which details how to calculate scattering loss due to snowflakes using Mie scattering theory \cite{ulaby2014scattering_book}.
Mie scattering theory applies to any spherical dielectric particle which can include snowflakes.
The dielectric constant of a snowflake, $\epsilon$, which is a mixture of water and ice, is calculated with Debye's mixture theory \cite{debye2} using the dielectric constants of water and ice at terahertz frequencies \cite{jiang2004ice}.
For propagation through air, the Mie extinction cross-section is defined as \cite{ulaby2014scattering_book}
\begin{equation}\label{eq:extinction}
    \sigma_{ext}(n, \chi) = \frac{2}{\chi^2} \sum^\infty_{m=1}(2m + 1)\Re\{a_m(n, \chi) + b_m(n, \chi)\},
\end{equation}
where $n = \epsilon^{1/2}$ and $\chi=\frac{2\pi r}{\lambda}$, and $r$ is the radius of the snowflake, $\lambda$ is the center frequency's wavelength, $a_m,\ b_m$ are the Mie coefficients, and $\Re$ indicates the real part of the coefficients.

For calculating the total attenuation due to snow scattering, $L_{snow}$, we use a modified form of the equation from \cite{ulaby2014scattering_book},
\begin{equation}\label{eq:snow_attenuation}
    L_{snow} = 4.343\cdot 10^3\ \sigma_{ext}\ N(r, R_{snow})\ r \ F\ ,
\end{equation}
where $N(r, R_{snow})$ is the Gunn-Marshall (G-M) snow size distribution \cite{gunn_marshall_snow}, $R_{snow}$ is the snowfall rate, and $F$ is the number of snowflakes present in the antenna beam.

The antenna beam's volume, $V$, is a cylinder with the height equal to the link distance, $H=70$~m, and the diameter equal to the lens width of the horn antenna, $D=0.122$~m.
The number of snowflakes in the antenna beam at any instance can be calculated based on the snow characteristics. For an estimated constant snow density, $\rho = 0.52$ g/cm$^3$, snowflake terminal velocity, $v = 1.5$ m/s, and snowflake weight, $m = 2.5$ mg \cite{mittleman2018snow}, the number of snowflakes in the antenna beam becomes a function of the time-varying snowfall rate, $R_{snow}(t)$, 
\begin{equation}\label{snowflake}
    F(t) = R_{snow}(t)
    \left( 
    \frac{\pi}{14.4}
    \frac{D^2 H \rho}{v m}
    \right)
    \ ,
\end{equation}
where $F(t)$ is the time-varying number of snowflakes from Eq.~\eqref{eq:snow_attenuation}.
The equation calculates the accumulation of snow in the antenna beam's cylindric volume after $\frac{D}{v}$ seconds scaled by the snow density, and it factors in unit conversion.
For $R_{snow}=0.45$ mm/hr (mean snow precipitation in Table \ref{tab:weather_stats}), there are approximately 15 snowflakes scattering the antenna beam.

\begin{figure}[t]
\centering
\includegraphics[width=1\linewidth]{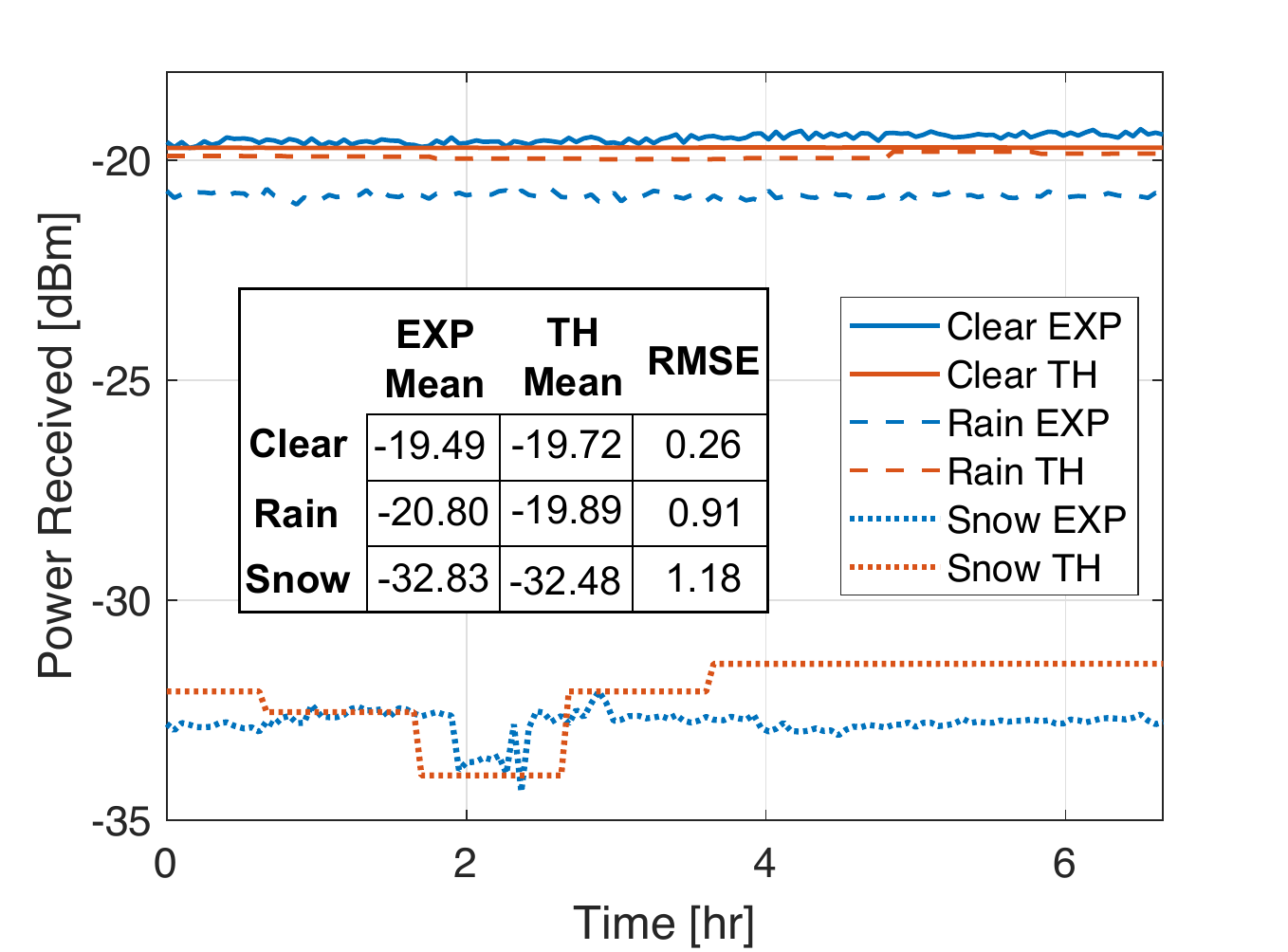}
\setlength\abovecaptionskip{-.3cm}
\setlength\belowcaptionskip{-.2cm}
\caption{Experimental and theoretical received power. Table: mean experimental and theoretical received power received and RMSE.
\vspace{-1em}
}
\label{fig:power_received}
\end{figure}

\subsubsection{Received Power Model}
\label{subsubsec:power_received_model}

The theoretical received power from Eq.~\eqref{eq:link_budget} is calculated with the modeled absorption and scattering losses.
The modeled $P_{r,T}$ is plotted as a function of time along with the experimentally measured received power, $P_{r,E}$, in Fig. \ref{fig:power_received}. We find that the experimental and modeled $P_r$ are relatively close. Their root mean square error (RMSE) is in the order of 1~dB for all the considered scenarios, as shown in the table inset in Fig.~\ref{fig:power_received}.

The clear and rain $P_{r,E}$ are mostly flat throughout their runtimes as the weather didn't significantly change during those days. The empirical link budget value matches the analytical one for clear conditions, but there is a constant offset of around 1~dB for the rain $P_{r,T}$. The latter is possibly due to a minor antenna misalignment.

The presented snow scattering model well matches the average received power measured experimentally. The model is also able to follow the trend at around the two hour mark in the plot. The snowfall greatly increased for one hour causing $P_{r,E}$ to drop by over 2~dB, and the developed model captures this effect.

\subsection{Capacity Analysis}
\label{subsec:Capacity_analysis}

Building on the link budget results, we analyze the theoretical capacity in Table~\ref{tab:capacity} focusing on three weather conditions. The table presents the mean \gls{snr}, Spectral Efficiency (SE), and capacity. We compare: (i)~the results for the idealistic hardware model (\gls{snr} is determined by the thermal noise) with (ii)~the results achievable with our measurement setup (\gls{snr} is also limited by the system noise floor). For the latter, the system noise power has been measured in the absence of any transmissions reaching the value of -51.3~dBm (common for all weather conditions).

\begin{table}[h!]
\begin{tabular}{|c|c|c|c|c|c|c|}
\cline{2-7}
\multicolumn{1}{l|}{}                & \multicolumn{3}{c|}{\textbf{Thermal noise}}                & \multicolumn{3}{c|}{\textbf{Thermal + System noise}}                \\
\cline{2-7}
\multicolumn{1}{l|}{}                & \textbf{SNR} & \textbf{SE} & \textbf{Cap.} & \textbf{SNR} & \textbf{SE} & \textbf{Capacity} \\ \cline{2-7}
\multicolumn{1}{c|}{\textbf{Clear}} & \multicolumn{1}{c|}{51.30}                             & \multicolumn{1}{c|}{17.04}                                          & \multicolumn{1}{c|}{340.83} & \multicolumn{1}{c|}{31.78} & \multicolumn{1}{c|}{10.56} & \multicolumn{1}{c|}{211.16}                               \\ \cline{2-7} 
\multicolumn{1}{c|}{\textbf{Rain}}  & \multicolumn{1}{c|}{49.99}                              & \multicolumn{1}{c|}{16.61}                                          & \multicolumn{1}{c|}{332.12}    & \multicolumn{1}{c|}{30.47}   & \multicolumn{1}{c|}{10.12} & \multicolumn{1}{c|}{202.46}                          \\ \cline{2-7} 
\multicolumn{1}{c|}{\textbf{Snow}}  & \multicolumn{1}{c|}{37.96}                             & \multicolumn{1}{c|}{12.61}                                          & \multicolumn{1}{c|}{252.20}     & \multicolumn{1}{c|}{18.44}    & \multicolumn{1}{c|}{6.15} & \multicolumn{1}{c|}{122.92}                        \\ \cline{2-7}
\end{tabular}
\setlength\abovecaptionskip{-.3cm}
\caption{Mean SNR (dB), spectral efficiency (SE, bit/s/Hz), and capacity (Gbit/s) of the link in different weather conditions.
}
\label{tab:capacity}
\end{table}

From Table~\ref{tab:capacity}, we observe that the difference between the link capacity for ``Clear'' and ``Rain'' is less than 10~Gbit/s or 5\%. Hence, the negative effect of light rain on wireless communications at 130--150~GHz is negligible from the SNR and the link capacity perspective. Meanwhile, the presence of snow drops the SNR at the receiver by over 13~dB leading to a capacity decrease of 90~Gbit/s. Hence, the decrease in capacity is 26\% for the idealistic hardware model and 42\% for our measurement setup. Still, the theoretical link capacity in the presence of snow is greater than 250~Gbit/s for the idealistic hardware model and as high as over 120~Gbit/s for our hardware. Notably higher capacity results are expected with the development of THz hardware featured by lower system noise.

\subsection{Multipath Profile}
\label{subsec:Multipath_profile}
\begin{figure}[t]
    \centering
    \includegraphics[width=1\linewidth]{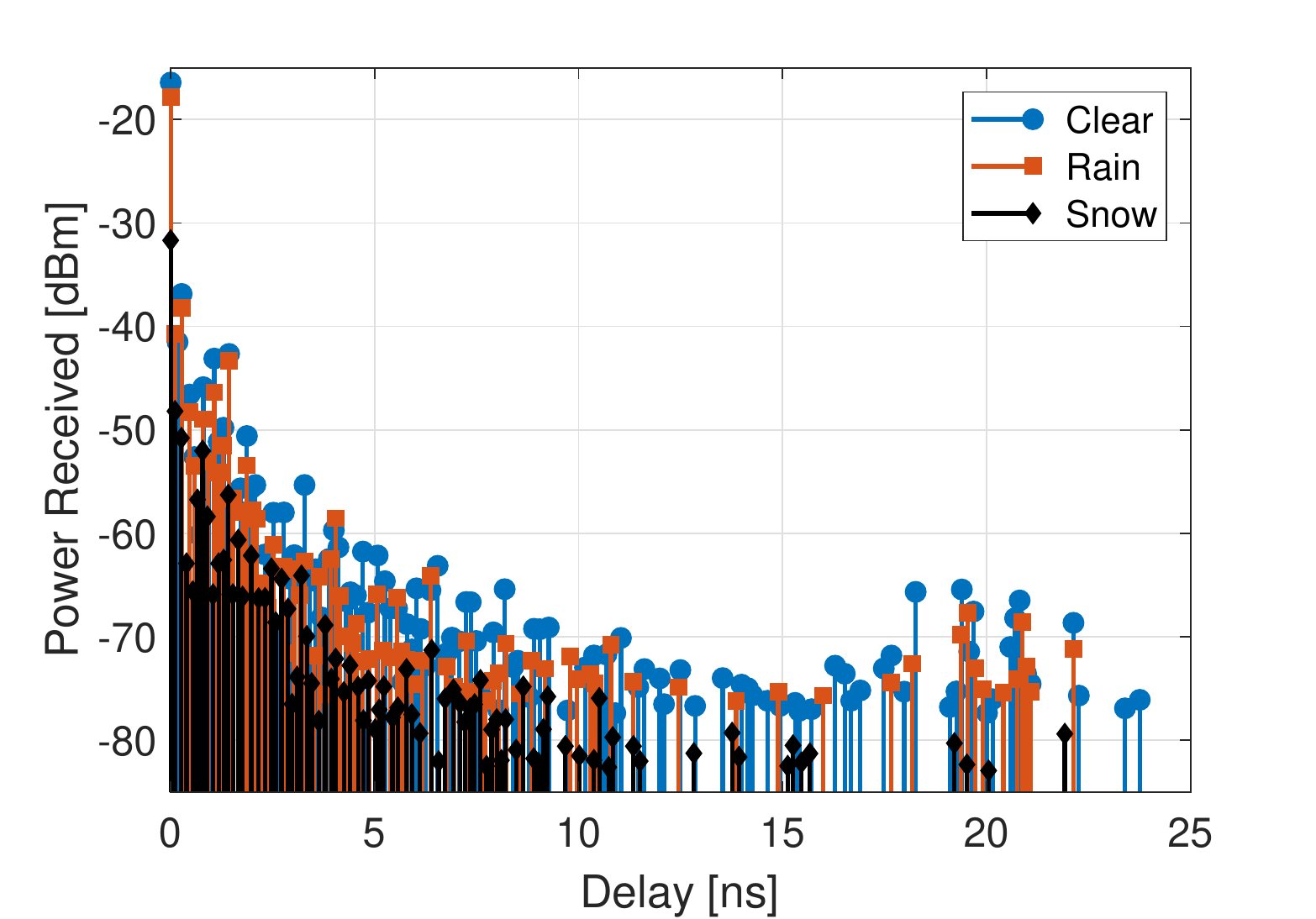}
    \setlength\abovecaptionskip{-.3cm}
    \setlength\belowcaptionskip{-.2cm}
    \caption{Multipath profiles for clear, rain, and snow weather.
    \vspace{-1em}
    }
    \label{fig:multipath_profiles}
\end{figure}

The sliding correlator generates multipath profiles by detecting delayed copies of the transmitted m-seq.
This measured channel impulse response and its multipath components (MPC) are plotted for one instance in Fig. \ref{fig:multipath_profiles}.
The channel impulses/peaks are estimated through global convergence method based on the time resolution and dynamic range of the system.

Although the system can detect delays up to 409.5~ns, we find that there are not a significant amount of MPCs beyond the 25~ns delay.
We notice that for clear and rain scenarios a cluster of MPCs with relatively high power is located around 20~ns. The same MPCs are 
weaker in the snow's channel impulse response.
This is likely due to a reflection off of the roof in front of the receiver, see Fig. \ref{fig:channel_sounder}~(b), whose power decreases when the roof is covered by a layer of snow. In the next paragraphs, we analyze the MPC profiles of the collected frames to estimate relevant channel metrics.

\subsection{Channel Metrics}
\label{subsec:metrics}
The outdoor point-to-point link is characterized in terms of the Rician K-factor and the root mean square (RMS) delay spread, which are among the crucial metrics to design wireless communication system.

\subsubsection{K-factor}
\label{subsubsec:k_factor}

The Rician K-factor is the ratio of the power in the LoS path ($P_{LoS}$) and other NLoS paths ($2~S^{2}$), and it thus given~by 
\begin{equation}\label{eq:k_factor}
K=\frac{P_{LoS}}{2~S^{2}}.
\end{equation}
In dB scale it is represented by $10log_{10}(K)$.
The K-factor is a significant component in static point-to-point links, as it characterises the distribution of the power across LoS and reflected paths.

The cumulative distribution function (CDF) of the K-factor for clear, rain, and snow, obtained from different time instances, is shown in Fig.~\ref{fig:k_factor}. The K-factor is reasonably steady for all three weather conditions, as we consider a static point-to-point link. Further, the metric variation is significantly low with the rain and snow rate changes.   
The K-factor in dB scale can be approximated by the normal distribution with mean, $\mu$, of 35.2 dB, 33.5 dB, and 29.8 dB, accompanied by the variance, $\sigma^{2}$, of 0.25, 0.19, and 0.22 for clear, rain, and snow, respectively. The statistical measurements are obtained by the maximum likelyhood estimation (MLE). Moreover, the K-factor decreases by 1.7~dB and 5.4~dB in case of rain and snow, respectively, compared to clear weather. 
The latter suggests that wireless THz communications and networks will need adaptive link modulation and coding scheme techniques to react to the different channel parameters with varying weather conditions.

\subsubsection{Delay Spread}
\label{subsubsec:Delay_spread}

\begin{figure}[t]
\centering
\includegraphics[width=1\linewidth]{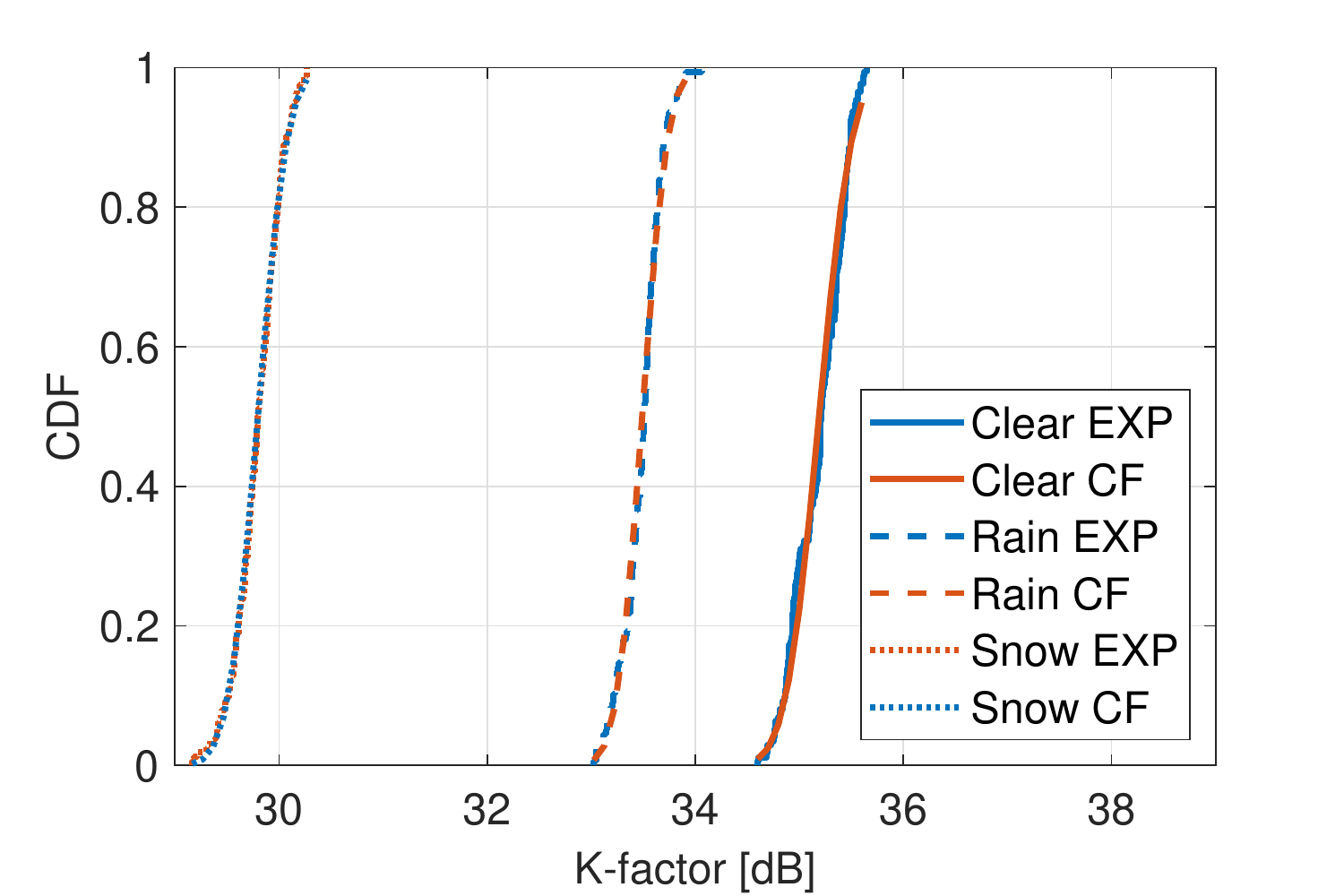}
\vskip -2ex
\caption{The CDF of K-factor [dB] in different weather conditions for experimental (EXP) results and curve fitting (CF) values.
\vspace{-1.5em}
}
\label{fig:k_factor}
\end{figure}

\begin{figure}[t]
\centering
\includegraphics[width=1\linewidth]{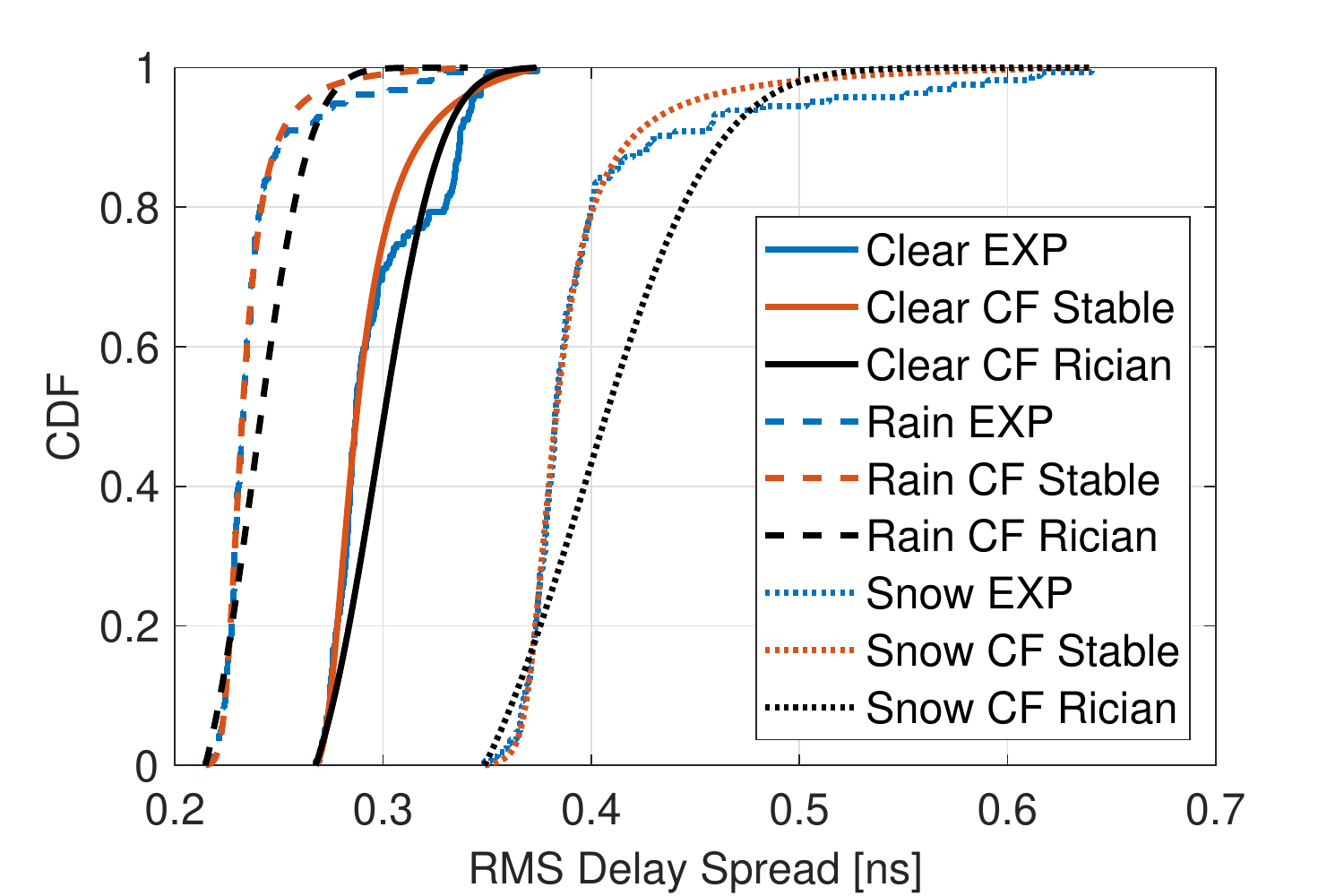}
\vskip -2ex
\caption{The CDF of RMS delay spread in different weather conditions for experimental (EXP) result and curve fitting (CF) value with the Stable and the Rician distributions.
\vspace{-1em}
}
\label{fig:d_spread}
\end{figure}

To describe the multipath richness of the link, the root mean square (RMS) delay spread, $\tau_{RMS}$, is calculated in different weather conditions as follows
\begin{equation}\label{eq:delay_spared}
\tau_{RMS} =\sqrt{\frac{\sum_{i}^{}(d_{i} - \hat{d})^2~p_{i}^{2}}{\sum_{i}^{}p_{i}^{2}}},
\end{equation}
where $d_{i}$ and $p_{i}$ are the delay and received power of propagation path $i$, respectively, and $\hat{d}$ is the mean delay represented by
\begin{equation}\label{eq:mean_delay}
\hat{d} =\frac{\sum_{i}^{}d_{i}~p_{i}}{\sum_{i}^{}p_{i}}.
\end{equation}
For our point-to-point link setup, the value for the delay spread is notably small (in the range of sub-nano~seconds). The CDF of the RMS delay spread (captured at different time instances) for three weather conditions is shown in Fig.~\ref{fig:d_spread}. As illustrated in Fig.~\ref{fig:d_spread}, the empirical CDF can be approximated by the CDF of the Rician distribution with the non-centrality values $\nu=0.29,0.23,0.39$, and the scale parameter values $\sigma=0.024,0.020,0.048$ for clear, rain, and snow conditions, respectively.

However, a closer fit is found with the stable distribution~\cite{nolan2020univariate_stable}, which is defined by a family of distributions capturing certain heavy-tail behaviors. In fact, for the point-to-point link with a high K-factor (as shown in Fig.~\ref{fig:k_factor}), the RMS delay spread exhibits different behaviors on the left and the right sides of the peak of its probability density function. Decaying on the left-hand side is significantly faster than on the right, which is observed from the CDF curves in Fig.~\ref{fig:d_spread}. Therefore, the stable distribution presents a closer fit for RMS delay spread, being able to model the complex nature of the tail and non-tail regions. The calculated distribution parameters are shown in Table~\ref{tab:CDF_stable}.

\begin{table}[h!]
\centering
\vspace{-.3cm}
\begin{tabular}{ccccc}
\multicolumn{1}{l}{}                & \textbf{\begin{tabular}[c]{@{}c@{}}$\alpha$\end{tabular}} & \textbf{\begin{tabular}[c]{@{}c@{}}$\beta$ \end{tabular}} & \textbf{\begin{tabular}[c]{@{}c@{}}$\gamma$\end{tabular}} & \textbf{\begin{tabular}[c]{@{}c@{}}$\delta$\end{tabular}} \\ \cline{2-5} 
\multicolumn{1}{c|}{\textbf{Clear}} & \multicolumn{1}{c|}{1.07}                             & \multicolumn{1}{c|}{1.0}                                          & \multicolumn{1}{c|}{0.0086}                                & \multicolumn{1}{c|}{0.28}                                          \\ \cline{2-5} 
\multicolumn{1}{c|}{\textbf{Rain}}  & \multicolumn{1}{c|}{1.28}                              & \multicolumn{1}{c|}{0.9}                                          & \multicolumn{1}{c|}{0.0052}                                & \multicolumn{1}{c|}{0.23}                                    \\ \cline{2-5} 
\multicolumn{1}{c|}{\textbf{Snow}}  & \multicolumn{1}{c|}{1.14}                             & \multicolumn{1}{c|}{0.8}                                          & \multicolumn{1}{c|}{0.0092}                                & \multicolumn{1}{c|}{0.38}                                    \\ \cline{2-5} 
\end{tabular}
\setlength\abovecaptionskip{-.3cm}
\caption{The $\alpha$ (first shape), $\beta$ (second shape), $\gamma$ (scale), and $\delta$ (location) parameters of the stable distribution for clear, rain, and snow.
}
\label{tab:CDF_stable}
\end{table}

Considering the 20~GHz bandwidth (symbol duration of 0.1~ns), the delay spread is large enough to create ISI. This aspect should be further considered in designing physical- and link-layer mechanisms for reliable communications. Further, the RMS delay spread value goes above 0.5~ns for more than 10\% of the cases in snow weather, which is due to the scattering of the signal by snowflakes. Therefore, we need to expand and/or adapt the state-of-the-art link adaptation techniques to still maintain reliable data exchange during harsh weather conditions.

\section{Conclusion}
\label{sec:conclusion}

In this paper, we reported the first channel sounding measurement results with snow and rain in the 130-150 GHz band. First, we described the channel sounder that was developed to support this effort, with details on the hardware and the signal processing techniques. Then, we analyzed the metrics related to the multipath profile, link budget, and several channel modeling parameters. The results show that adverse weather conditions introduce additional impairments in the sub-terahertz channel, however, the measured setup can still achieve attractive SNR and link capacity.

For future work, we plan to gather data in additional scenarios (indoor, outdoor, aerial) with different structural and geometrical aspects, blockage, antenna gain, and atmospheric conditions at different frequencies above 100 GHz. Specifically, we plan to measure the probability distribution as well as the correlation of the metrics, such as delay spread, angular spread, and path loss coefficient, which will contribute to creation of efficient and reliable communication links. Further, we will look into the spectral efficient schemes in terms of modulation~\cite{sen2022optimized_apsk} design and THz multiple-input multiple-output (MIMO) channels~\cite{akyildiz2016realizing}. 
Finding the characteristics for MIMO and single-input single-output channels will provide the necessary tools for future high-speed wireless communications above 100~GHz.

\vspace{-.2cm}
\begin{acks}
This work was partially funded by U.S. National Science Foundation under grants CNS-2011411 and CNS-2120447, U.S. Air Force Research Laboratory grant FA8750-20-1-0200, and OUSD(R\&E) through U.S. Army Research Laboratory cooperative agreement number W911NF-19-2-0221. 
The authors would like to thank the Northeastern University personnel (Joe Lalley, Beshad Moghaddam, Elham Ghabbour, and Patrick Connolly) who helped with the safety/logistics of the channel sounding campaign.
\end{acks}

\bibliographystyle{ACM-Reference-Format.bst}
\bibliography{bibl.bib}


\begin{thebibliography}{25}


\ifx \showCODEN    \undefined \def \showCODEN     #1{\unskip}     \fi
\ifx \showDOI      \undefined \def \showDOI       #1{#1}\fi
\ifx \showISBNx    \undefined \def \showISBNx     #1{\unskip}     \fi
\ifx \showISBNxiii \undefined \def \showISBNxiii  #1{\unskip}     \fi
\ifx \showISSN     \undefined \def \showISSN      #1{\unskip}     \fi
\ifx \showLCCN     \undefined \def \showLCCN      #1{\unskip}     \fi
\ifx \shownote     \undefined \def \shownote      #1{#1}          \fi
\ifx \showarticletitle \undefined \def \showarticletitle #1{#1}   \fi
\ifx \showURL      \undefined \def \showURL       {\relax}        \fi
\providecommand\bibfield[2]{#2}
\providecommand\bibinfo[2]{#2}
\providecommand\natexlab[1]{#1}
\providecommand\showeprint[2][]{arXiv:#2}

\bibitem[\protect\citeauthoryear{Akyildiz, Han, Hu, Nie, and Jornet}{Akyildiz
  et~al\mbox{.}}{2022}]%
        {akyildiz2022terahertz}
\bibfield{author}{\bibinfo{person}{Ian~F. Akyildiz}, \bibinfo{person}{Chong
  Han}, \bibinfo{person}{Zhifeng Hu}, \bibinfo{person}{Shuai Nie}, {and}
  \bibinfo{person}{Josep~Miquel Jornet}.} \bibinfo{year}{2022}\natexlab{}.
\newblock \showarticletitle{Terahertz Band Communication: An Old Problem
  Revisited and Research Directions for the Next Decade}.
\newblock \bibinfo{journal}{\emph{IEEE Transactions on Communications}}
  \bibinfo{volume}{70}, \bibinfo{number}{6} (\bibinfo{date}{June}
  \bibinfo{year}{2022}), \bibinfo{pages}{4250--4285}.
\newblock
\showISSN{1558-0857}
\urldef\tempurl%
\url{https://doi.org/10.1109/TCOMM.2022.3171800}
\showDOI{\tempurl}


\bibitem[\protect\citeauthoryear{Akyildiz and Jornet}{Akyildiz and
  Jornet}{2016}]%
        {akyildiz2016realizing}
\bibfield{author}{\bibinfo{person}{Ian~F Akyildiz} {and}
  \bibinfo{person}{Josep~Miquel Jornet}.} \bibinfo{year}{2016}\natexlab{}.
\newblock \showarticletitle{Realizing ultra-massive MIMO (1024$\times$ 1024)
  communication in the (0.06--10) terahertz band}.
\newblock \bibinfo{journal}{\emph{Nano Communication Networks}}
  \bibinfo{volume}{8} (\bibinfo{year}{2016}), \bibinfo{pages}{46--54}.
\newblock


\bibitem[\protect\citeauthoryear{Chen, Li, Han, Yu, and Wang}{Chen
  et~al\mbox{.}}{2021}]%
        {chen2021channel}
\bibfield{author}{\bibinfo{person}{Yi Chen}, \bibinfo{person}{Yuanbo Li},
  \bibinfo{person}{Chong Han}, \bibinfo{person}{Ziming Yu}, {and}
  \bibinfo{person}{Guangjian Wang}.} \bibinfo{year}{2021}\natexlab{}.
\newblock \showarticletitle{{Channel Measurement and Ray-Tracing-Statistical
  Hybrid Modeling for Low-Terahertz Indoor Communications}}.
\newblock \bibinfo{journal}{\emph{IEEE Transactions on Wireless
  Communications}}  \bibinfo{volume}{,Early Access} (\bibinfo{year}{2021}).
\newblock
\urldef\tempurl%
\url{https://doi.org/10.1109/TWC.2021.3090781}
\showDOI{\tempurl}


\bibitem[\protect\citeauthoryear{Giordani, Polese, Mezzavilla, Rangan, and
  Zorzi}{Giordani et~al\mbox{.}}{2020}]%
        {giordani2020toward}
\bibfield{author}{\bibinfo{person}{Marco Giordani}, \bibinfo{person}{Michele
  Polese}, \bibinfo{person}{Marco Mezzavilla}, \bibinfo{person}{Sundeep
  Rangan}, {and} \bibinfo{person}{Michele Zorzi}.}
  \bibinfo{year}{2020}\natexlab{}.
\newblock \showarticletitle{{Toward 6G Networks: Use Cases and Technologies}}.
\newblock \bibinfo{journal}{\emph{IEEE Communications Magazine}}
  \bibinfo{volume}{58}, \bibinfo{number}{3} (\bibinfo{date}{March}
  \bibinfo{year}{2020}), \bibinfo{pages}{55--61}.
\newblock
\showISSN{1558-1896}
\urldef\tempurl%
\url{https://doi.org/10.1109/MCOM.001.1900411}
\showDOI{\tempurl}


\bibitem[\protect\citeauthoryear{{Gunn} and {Marshall}}{{Gunn} and
  {Marshall}}{1958}]%
        {gunn_marshall_snow}
\bibfield{author}{\bibinfo{person}{K.~L.~S. {Gunn}} {and}
  \bibinfo{person}{J.~S. {Marshall}}.} \bibinfo{year}{1958}\natexlab{}.
\newblock \showarticletitle{{The Distribution with Size of Aggregate
  Snowflakes.}}
\newblock \bibinfo{journal}{\emph{Journal of Atmospheric Sciences}}
  \bibinfo{volume}{15}, \bibinfo{number}{5} (\bibinfo{date}{Oct.}
  \bibinfo{year}{1958}), \bibinfo{pages}{452--461}.
\newblock


\bibitem[\protect\citeauthoryear{{International Telecommunication
  Union}}{{International Telecommunication Union}}{2005}]%
        {iturp838}
\bibfield{author}{\bibinfo{person}{{International Telecommunication Union}}.}
  \bibinfo{year}{2005}\natexlab{}.
\newblock \bibinfo{title}{Recommendation ITU-R P.838-3: Specific attenuation
  model for rain for use in prediction methods}.
\newblock
\newblock


\bibitem[\protect\citeauthoryear{{International Telecommunication
  Union}}{{International Telecommunication Union}}{2019}]%
        {iturp676}
\bibfield{author}{\bibinfo{person}{{International Telecommunication Union}}.}
  \bibinfo{year}{2019}\natexlab{}.
\newblock \bibinfo{title}{Recommendation ITU-R P.676-12: Attenuation by
  atmospheric gases and related effects}.
\newblock
\newblock


\bibitem[\protect\citeauthoryear{Jiang and Wu}{Jiang and Wu}{2004}]%
        {jiang2004ice}
\bibfield{author}{\bibinfo{person}{Jonathan~H Jiang} {and}
  \bibinfo{person}{Dong~L Wu}.} \bibinfo{year}{2004}\natexlab{}.
\newblock \showarticletitle{Ice and water permittivities for millimeter and
  sub-millimeter remote sensing applications}.
\newblock \bibinfo{journal}{\emph{Atmospheric Science Letters}}
  \bibinfo{volume}{5}, \bibinfo{number}{7} (\bibinfo{date}{Oct.}
  \bibinfo{year}{2004}), \bibinfo{pages}{146--151}.
\newblock
\showISSN{1530-261X, 1530-261X}
\urldef\tempurl%
\url{https://doi.org/10.1002/asl.77}
\showDOI{\tempurl}


\bibitem[\protect\citeauthoryear{Jornet and Akyildiz}{Jornet and
  Akyildiz}{2011}]%
        {jornet2011channel}
\bibfield{author}{\bibinfo{person}{Josep~Miquel Jornet} {and}
  \bibinfo{person}{Ian~F Akyildiz}.} \bibinfo{year}{2011}\natexlab{}.
\newblock \showarticletitle{Channel modeling and capacity analysis for
  electromagnetic wireless nanonetworks in the terahertz band}.
\newblock \bibinfo{journal}{\emph{IEEE Transactions on Wireless
  Communications}} \bibinfo{volume}{10}, \bibinfo{number}{10}
  (\bibinfo{date}{Aug} \bibinfo{year}{2011}), \bibinfo{pages}{3211--3221}.
\newblock


\bibitem[\protect\citeauthoryear{Jornet, Sen, and Ariyarathna}{Jornet
  et~al\mbox{.}}{2022}]%
        {jornet2022ultra}
\bibfield{author}{\bibinfo{person}{Josep~Miquel Jornet},
  \bibinfo{person}{Priyangshu Sen}, {and} \bibinfo{person}{Viduneth
  Ariyarathna}.} \bibinfo{year}{2022}\natexlab{}.
\newblock \showarticletitle{Ultra-Broadband Networking Systems Testbed at
  Northeastern University}.
\newblock In \bibinfo{booktitle}{\emph{THz Communications}}.
  \bibinfo{publisher}{Springer}, \bibinfo{pages}{473--476}.
\newblock


\bibitem[\protect\citeauthoryear{Ju, Xing, Kanhere, and Rappaport}{Ju
  et~al\mbox{.}}{2021}]%
        {Statchannel}
\bibfield{author}{\bibinfo{person}{Shihao Ju}, \bibinfo{person}{Yunchou Xing},
  \bibinfo{person}{Ojas Kanhere}, {and} \bibinfo{person}{Theodore~S.
  Rappaport}.} \bibinfo{year}{2021}\natexlab{}.
\newblock \showarticletitle{Millimeter Wave and Sub-Terahertz Spatial
  Statistical Channel Model for an Indoor Office Building}.
\newblock \bibinfo{journal}{\emph{IEEE Journal on Selected Areas in
  Communications}} \bibinfo{volume}{39}, \bibinfo{number}{6}
  (\bibinfo{date}{Apr} \bibinfo{year}{2021}), \bibinfo{pages}{1561--1575}.
\newblock


\bibitem[\protect\citeauthoryear{Ma, Adelberg, Shrestha, Moeller, and
  Mittleman}{Ma et~al\mbox{.}}{2018}]%
        {mittleman2018snow}
\bibfield{author}{\bibinfo{person}{Jianjun Ma}, \bibinfo{person}{Jacob
  Adelberg}, \bibinfo{person}{Rabi Shrestha}, \bibinfo{person}{Lothar Moeller},
  {and} \bibinfo{person}{Daniel Mittleman}.} \bibinfo{year}{2018}\natexlab{}.
\newblock \showarticletitle{The Effect of Snow on a Terahertz Wireless Data
  Link}.
\newblock \bibinfo{journal}{\emph{Journal of infrared, millimeter and terahertz
  waves}}  \bibinfo{volume}{39} (\bibinfo{date}{03} \bibinfo{year}{2018}).
\newblock
\urldef\tempurl%
\url{https://doi.org/10.1007/s10762-018-0486-2}
\showDOI{\tempurl}


\bibitem[\protect\citeauthoryear{Moltchanov, Gaidamaka, Ostrikova, Beschastnyi,
  Koucheryavy, and Samouylov}{Moltchanov et~al\mbox{.}}{2022}]%
        {moltchanov2022blockage}
\bibfield{author}{\bibinfo{person}{Dmitri Moltchanov}, \bibinfo{person}{Yuliya
  Gaidamaka}, \bibinfo{person}{Darya Ostrikova}, \bibinfo{person}{Vitalii
  Beschastnyi}, \bibinfo{person}{Yevgeni Koucheryavy}, {and}
  \bibinfo{person}{Konstantin Samouylov}.} \bibinfo{year}{2022}\natexlab{}.
\newblock \showarticletitle{Ergodic Outage and Capacity of Terahertz Systems
  Under Micromobility and Blockage Impairments}.
\newblock \bibinfo{journal}{\emph{IEEE Transactions on Wireless
  Communications}} \bibinfo{volume}{21}, \bibinfo{number}{5}
  (\bibinfo{date}{May} \bibinfo{year}{2022}), \bibinfo{pages}{3024--3039}.
\newblock
\showISSN{1558-2248}
\urldef\tempurl%
\url{https://doi.org/10.1109/TWC.2021.3117583}
\showDOI{\tempurl}


\bibitem[\protect\citeauthoryear{Moon, Jeon, and Grischkowsky}{Moon
  et~al\mbox{.}}{2015}]%
        {moon2015long}
\bibfield{author}{\bibinfo{person}{Eom-Bae Moon}, \bibinfo{person}{Tae-In
  Jeon}, {and} \bibinfo{person}{Daniel~R. Grischkowsky}.}
  \bibinfo{year}{2015}\natexlab{}.
\newblock \showarticletitle{{Long-Path THz-TDS Atmospheric Measurements Between
  Buildings}}.
\newblock \bibinfo{journal}{\emph{IEEE Transactions on Terahertz Science and
  Technology}} \bibinfo{volume}{5}, \bibinfo{number}{5} (\bibinfo{date}{Sep.}
  \bibinfo{year}{2015}), \bibinfo{pages}{742--750}.
\newblock
\showISSN{2156-3446}
\urldef\tempurl%
\url{https://doi.org/10.1109/TTHZ.2015.2443491}
\showDOI{\tempurl}


\bibitem[\protect\citeauthoryear{Nolan}{Nolan}{2020}]%
        {nolan2020univariate_stable}
\bibfield{author}{\bibinfo{person}{John~P Nolan}.}
  \bibinfo{year}{2020}\natexlab{}.
\newblock \bibinfo{booktitle}{\emph{Univariate stable distributions}}.
\newblock \bibinfo{publisher}{Springer}.
\newblock


\bibitem[\protect\citeauthoryear{Oguchi}{Oguchi}{1983}]%
        {debye2}
\bibfield{author}{\bibinfo{person}{T. Oguchi}.}
  \bibinfo{year}{1983}\natexlab{}.
\newblock \showarticletitle{Electromagnetic wave propagation and scattering in
  rain and other hydrometeors}.
\newblock \bibinfo{journal}{\emph{Proc. IEEE}} \bibinfo{volume}{71},
  \bibinfo{number}{9} (\bibinfo{year}{1983}), \bibinfo{pages}{1029--1078}.
\newblock
\urldef\tempurl%
\url{https://doi.org/10.1109/PROC.1983.12724}
\showDOI{\tempurl}


\bibitem[\protect\citeauthoryear{Petrov, Moltchanov, Koucheryavy, and
  Jornet}{Petrov et~al\mbox{.}}{2020}]%
        {petrov2020micro}
\bibfield{author}{\bibinfo{person}{Vitaly Petrov}, \bibinfo{person}{Dmitri
  Moltchanov}, \bibinfo{person}{Yevgeni Koucheryavy}, {and}
  \bibinfo{person}{Josep~M. Jornet}.} \bibinfo{year}{2020}\natexlab{}.
\newblock \showarticletitle{Capacity and Outage of Terahertz Communications
  With User Micro-Mobility and Beam Misalignment}.
\newblock \bibinfo{journal}{\emph{IEEE Transactions on Vehicular Technology}}
  \bibinfo{volume}{69}, \bibinfo{number}{6} (\bibinfo{year}{2020}),
  \bibinfo{pages}{6822--6827}.
\newblock
\urldef\tempurl%
\url{https://doi.org/10.1109/TVT.2020.2988600}
\showDOI{\tempurl}


\bibitem[\protect\citeauthoryear{Polese, Ariyarathna, Sen, Siles, Restuccia,
  Melodia, and Jornet}{Polese et~al\mbox{.}}{2022}]%
        {polese2022dynamic}
\bibfield{author}{\bibinfo{person}{Michele Polese}, \bibinfo{person}{Viduneth
  Ariyarathna}, \bibinfo{person}{Priyangshu Sen}, \bibinfo{person}{Jose~V
  Siles}, \bibinfo{person}{Francesco Restuccia}, \bibinfo{person}{Tommaso
  Melodia}, {and} \bibinfo{person}{Josep~M Jornet}.}
  \bibinfo{year}{2022}\natexlab{}.
\newblock \showarticletitle{Dynamic spectrum sharing between active and passive
  users above 100 GHz}.
\newblock \bibinfo{journal}{\emph{Communications Engineering}}
  \bibinfo{volume}{1}, \bibinfo{number}{1} (\bibinfo{year}{2022}),
  \bibinfo{pages}{1--9}.
\newblock


\bibitem[\protect\citeauthoryear{Polese, Jornet, Melodia, and Zorzi}{Polese
  et~al\mbox{.}}{2020}]%
        {polese2020toward}
\bibfield{author}{\bibinfo{person}{Michele Polese},
  \bibinfo{person}{Josep~Miquel Jornet}, \bibinfo{person}{Tommaso Melodia},
  {and} \bibinfo{person}{Michele Zorzi}.} \bibinfo{year}{2020}\natexlab{}.
\newblock \showarticletitle{{Toward End-to-End, Full-Stack 6G Terahertz
  Networks}}.
\newblock \bibinfo{journal}{\emph{IEEE Communications Magazine}}
  \bibinfo{volume}{58}, \bibinfo{number}{11} (\bibinfo{date}{November}
  \bibinfo{year}{2020}), \bibinfo{pages}{48--54}.
\newblock
\showISSN{1558-1896}


\bibitem[\protect\citeauthoryear{Redmill and Bertin}{Redmill and
  Bertin}{2021}]%
        {redmill20216g}
\bibfield{author}{\bibinfo{person}{Guy Redmill} {and} \bibinfo{person}{Emmanuel
  Bertin}.} \bibinfo{year}{2021}\natexlab{}.
\newblock \showarticletitle{{6G: The Path Toward Standardization}}.
\newblock \bibinfo{journal}{\emph{Shaping Future 6G Networks: Needs, Impacts,
  and Technologies}} (\bibinfo{year}{2021}), \bibinfo{pages}{23--37}.
\newblock


\bibitem[\protect\citeauthoryear{Renaud and Federici}{Renaud and
  Federici}{2019}]%
        {federici2019snow}
\bibfield{author}{\bibinfo{person}{{Dylan L.} Renaud} {and}
  \bibinfo{person}{{John F.} Federici}.} \bibinfo{year}{2019}\natexlab{}.
\newblock \showarticletitle{Terahertz Attenuation in Snow and Sleet}.
\newblock \bibinfo{journal}{\emph{Journal of Infrared, Millimeter, and
  Terahertz Waves}} \bibinfo{volume}{40}, \bibinfo{number}{8}
  (\bibinfo{date}{15 Aug.} \bibinfo{year}{2019}), \bibinfo{pages}{868--877}.
\newblock
\showISSN{1866-6892}
\urldef\tempurl%
\url{https://doi.org/10.1007/s10762-019-00607-y}
\showDOI{\tempurl}


\bibitem[\protect\citeauthoryear{Sen, Ariyarathna, and Jornet}{Sen
  et~al\mbox{.}}{2022}]%
        {sen2022optimized_apsk}
\bibfield{author}{\bibinfo{person}{Priyangshu Sen}, \bibinfo{person}{Viduneth
  Ariyarathna}, {and} \bibinfo{person}{Josep~M Jornet}.}
  \bibinfo{year}{2022}\natexlab{}.
\newblock \showarticletitle{An Optimized M-ary Amplitude Phase Shift Keying
  Scheme for Ultrabroadband Terahertz Communication}. In
  \bibinfo{booktitle}{\emph{2022 IEEE 19th Annual Consumer Communications \&
  Networking Conference (CCNC)}}. IEEE, \bibinfo{pages}{661--666}.
\newblock


\bibitem[\protect\citeauthoryear{Sen, Ariyarathna, Madanayake, and Jornet}{Sen
  et~al\mbox{.}}{2021}]%
        {sen2021versatile}
\bibfield{author}{\bibinfo{person}{Priyangshu Sen}, \bibinfo{person}{Viduneth
  Ariyarathna}, \bibinfo{person}{Arjuna Madanayake}, {and}
  \bibinfo{person}{Josep~M Jornet}.} \bibinfo{year}{2021}\natexlab{}.
\newblock \showarticletitle{A versatile experimental testbed for ultrabroadband
  communication networks above 100 GHz}.
\newblock \bibinfo{journal}{\emph{Computer Networks}}  \bibinfo{volume}{193}
  (\bibinfo{year}{2021}), \bibinfo{pages}{108092}.
\newblock


\bibitem[\protect\citeauthoryear{Singh, Andrello, Thawdar, and Jornet}{Singh
  et~al\mbox{.}}{2020}]%
        {singh2020design}
\bibfield{author}{\bibinfo{person}{Arjun Singh}, \bibinfo{person}{Michael
  Andrello}, \bibinfo{person}{Ngwe Thawdar}, {and}
  \bibinfo{person}{Josep~Miquel Jornet}.} \bibinfo{year}{2020}\natexlab{}.
\newblock \showarticletitle{Design and Operation of a Graphene-Based Plasmonic
  Nano-Antenna Array for Communication in the Terahertz Band}.
\newblock \bibinfo{journal}{\emph{IEEE Journal on Selected Areas in
  Communications}} \bibinfo{volume}{38}, \bibinfo{number}{9}
  (\bibinfo{date}{Sep.} \bibinfo{year}{2020}), \bibinfo{pages}{2104--2117}.
\newblock
\showISSN{1558-0008}
\urldef\tempurl%
\url{https://doi.org/10.1109/JSAC.2020.3000881}
\showDOI{\tempurl}


\bibitem[\protect\citeauthoryear{Ulaby, Long, Blackwell, Elachi, Fung, Ruf,
  Sarabandi, Zyl, and Zebker}{Ulaby et~al\mbox{.}}{2014}]%
        {ulaby2014scattering_book}
\bibfield{author}{\bibinfo{person}{Fawwaz Ulaby}, \bibinfo{person}{David Long},
  \bibinfo{person}{William Blackwell}, \bibinfo{person}{Charles Elachi},
  \bibinfo{person}{Adrian Fung}, \bibinfo{person}{Christopher Ruf},
  \bibinfo{person}{K. Sarabandi}, \bibinfo{person}{Jakob Zyl}, {and}
  \bibinfo{person}{Howard Zebker}.} \bibinfo{year}{2014}\natexlab{}.
\newblock \bibinfo{booktitle}{\emph{Microwave Radar and Radiometric Remote
  Sensing}}.
\newblock
\showISBNx{978-0-472-11935-6}


\end{thebibliography}

\end{document}